\documentclass[prl,twocolumn,aps,superscriptaddress,preprintnumbers,letterpaper,showpacs]{revtex4}
\usepackage{amsmath,amssymb}
\usepackage{epsfig}
\usepackage{graphicx}
\usepackage{amsmath}
\usepackage{amsfonts}
\usepackage{epstopdf}

\begin{document}
\renewcommand{\vec}{\boldsymbol}
\newcommand{\mc}{M_{\mathrm{c}}}
\newcommand{\mn}{M_{\mathrm{n}}}
\newcommand{\mnc}{M_{\mathrm{nc}}} 
\newcommand{\mr}{M_{\mathrm{R}}}
\newcommand{\gma}{\gamma}
\newcommand{\gmat}{\tilde{\gamma}}
\newcommand{\pc}{\vec{p}_{c}}
\newcommand{\pn}{\vec{p}_{n}}
\newcommand{\bes}{{}^{7}\mathrm{Be}}
\newcommand{\besst}{{}^{7}\mathrm{Be}^{\ast}}
\newcommand{\be}{{}^{8}\mathrm{B}}
\renewcommand{\S}[2]{{}^{#1}S_{#2}}
\renewcommand{\P}[2]{{}^{#1}P_{#2}}
\newcommand{\gone}{g_{(\S{3}{1})}}
\newcommand{\gtwo}{g_{(\S{5}{2})}}
\newcommand{\gthree}{g_{(\S{3}{1}^{*})}}
\newcommand{\aone}{a_{(\S{3}{1})}}
\newcommand{\atwo}{\ensuremath{a_{(\S{5}{2})}}}
\newcommand{\hone}{h_{(\P{3}{2})}}
\newcommand{\htwo}{h_{(\P{5}{2})}}
\newcommand{\hpt}{h_{t}}
\newcommand{\hthree}{h_{(\P{3}{2}^{*})}}
\newcommand{\honet}{\tilde{h}_{(\P{3}{1})}}
\newcommand{\htwot}{\tilde{h}_{(\P{5}{1})}}
\newcommand{\Xone}{{X}_{(\S{3}{1})}}
\newcommand{\Xtwo}{{X}_{(\S{5}{2})}}
\newcommand{\Xonet}{\tilde{X}_{(\S{3}{1})}}
\newcommand{\Xtwot}{\tilde{X}_{(\S{5}{2})}}
\newcommand{\V}[1]{\vec{V}_{#1}}
\newcommand{\fdu}[2]{{#1}^{\dagger #2}}
\newcommand{\fdd}[2]{{#1}^{\dagger}_{#2}}
\newcommand{\fu}[2]{{#1}^{#2}}
\newcommand{\fd}[2]{{#1}_{#2}}
\newcommand{\T}[2]{T_{#1}^{\, #2}}
\newcommand{\e}{\vec{\epsilon}}
\newcommand{\es}{\e^{*}}
\newcommand{\cw}[2]{\chi^{(#2)}_{#1}}
\newcommand{\cwc}[2]{\chi^{(#2)*}_{#1}}
\newcommand{\cwf}[1]{F_{#1}}
\newcommand{\cwg}[1]{G_{#1}}
\newcommand{\ke}{k}
\newcommand{\kest}{k_{\ast}}
\newcommand{\kc}{k_{C}}
\newcommand{\upartial}[1]{\partial^{#1}}
\newcommand{\dpartial}[1]{\partial_{#1}}
\newcommand{\etae}{\eta_{E}}
\newcommand{\etab}{\eta_{B}}
\newcommand{\etaest}{\eta_{E\ast}}
\newcommand{\etabst}{\eta_{B\ast}}
\newcommand{\vecpt}[1]{\hat{\vec{#1}}}
\newcommand{\uY}[2]{Y_{#1}^{#2}}
\newcommand{\dY}[2]{Y_{#1 #2}}

\title{Marrying {\it ab initio} calculations and Halo EFT: the case of ${}^7{\rm Be} + p \rightarrow {}^8{\rm B} + \gamma$}


%
\author{Xilin Zhang} 
\author{Kenneth M.~Nollett} 
\author{D.~R.~Phillips} 
\affiliation{Institute of Nuclear and Particle Physics and Department of
Physics and Astronomy, Ohio University, Athens, OH\ \ 45701, USA}

\date{December 2013}

\begin{abstract}
We report a leading-order (LO) calculation of $^7\mathrm{Be}(p,\gamma)^8\mathrm{B}$ in a low-energy effective field theory. $\be$ is treated as a shallow proton$+\bes$ core and proton$+\bes^{*}$ (core excitation) $p$-wave bound state. The couplings are fixed using measured binding energies and proton-$\bes$ $s$-wave scattering lengths, together with $\be$ asymptotic normalization coefficients from {\it ab initio} calculations. We obtain a zero-energy $S$-factor of $18.2 \pm 1.2~({\rm ANC~only})$ eV b. Given that this is a LO result it is consistent with the recommended value $S(0)=20.8\pm1.6$  eV b. Our computed $S(E)$ compares favorably with experimental data on  $^7\mathrm{Be}(p,\gamma)^8\mathrm{B}$ for $E <0.4$ MeV. We emphasize the important role of proton-$\bes$ scattering parameters in determining the energy dependence of $S(E)$, and demonstrate that their present uncertainties significantly limit attempts to extrapolate these data  to stellar energies.
\end{abstract}


\maketitle

{\em Introduction.---}
The cross section of the reaction
$^7\mathrm{Be}(p,\gamma)^8\mathrm{B}$ is important for constraining
properties of neutrino oscillations and solar composition through solar-neutrino experiments (e.g.,
\cite{Adelberger:2010qa,Haxton:2013}).  It must be known at very low ($\sim 20$
keV) energies, and this presents a general problem common to nearly
all processes in stellar nuclear burning \cite{rolfs-rodney}.  Such
cross sections are very small at these energies because of Coulomb
barriers.  Most direct measurements have to be carried out at higher
energies with larger cross sections and extrapolated down to stellar
energies using models.

The accuracy of this approach often suffers from a scarcity of
ancillary constraints like elastic cross sections. It also depends
on model assumptions regarding, e.g., the number of $R$-matrix poles
or the shape of a potential well. Often the relevance of the constraints to their desired application is doubtful but hard to quantify (as reviewed for
$^7\mathrm{Be}(p,\gamma)^8\mathrm{B}$ in
Ref.~\cite{Adelberger:2010qa}).  Recent advances in \textit{ab initio}
calculations hold promise to ameliorate the situation
\cite{Navratil:2011sa}, but accurate and complete three-nucleon forces
are yet to be fully incorporated there~\cite{Hupin:2013wsa}.

The degrees of freedom and general philosophy of effective field
theory (EFT) treatments of such reactions are the same as those in
$R$-matrix and potential-model calculations.  EFTs also fix their
couplings from particular observables and then predict others. But the
EFT's systematic expansion provides a quantitative estimate of the
uncertainty at each order of approximation and a framework to improve
accuracy. The particular EFT we use here, Halo
EFT~\cite{vK99,Ka98A,Ka98B,Be02,Bd03,Hammer:2011ye,Rupak:2011nk,Canham:2008,Higa:2008},
does this by exploiting the separation of energy scales associated
with the presence of clusters.  Much of a nucleon-level reaction
calculation amounts to an indirect treatment of the collective motion
of clusters through and around Coulomb and centrifugal barriers, so
Halo EFT can profitably be combined with {\it ab initio}
calculations. We extract asymptotic normalization coefficients (ANCs)
from {\it ab initio} eight-body
calculations~\cite{Nollett:2011qf,Navratil:2011sa}, combine them with
$p$-$\bes$ scattering lengths from both experiment \cite{Angulo:2003}
and theory \cite{Navratil:2011sa}, and perform a predictive Halo EFT
computation of $^7\mathrm{Be}(p,\gamma)^8\mathrm{B}$. This extends
Ref.~\cite{Ryberg:2013}'s calculation of E1 proton capture to the case
of a $p$-wave bound state.

\begin{table} [h!]
  \begin{tabular}{|c|c|c|}
 \hline
 Momentum scale & Definition & Value\\
 \hline
  $\kc\sim \gamma$ & $Q_c Q_n \alpha_{EM} \mr$ & 24.02 MeV\\
 
 $\gamma$ & $\sqrt{2 M_R B_{\be}}$ & 15.04 MeV\\
 
  $\Lambda$ & $\sqrt{2 M'_R B_{\bes}}$ & 70 MeV\\
 
 $\gamma^* \sim \gamma$ & $\sqrt{2 M_R(B_{\be} + E^*)}$ & 30.53 MeV\\
 
 $\gamma_\Delta\sim \gamma$ & $\sqrt{2 M_R E^*}$ & 26.57 MeV\\

$a_{\S{3}{1}}$, $a_{\S{5}{2}}$ $\sim 1/\gamma$ & scattering lengths & Varies\\
 
 $r_0$ $\sim 1/\Lambda$ & $l=0$ effective ranges & Varies\\
 
 $a_1 \sim \gamma^{-2}\Lambda^{-1}$ &  scattering volume & 1054.1 $\mathrm{fm}^{3}$ \\
 
 $r_1 \sim \Lambda$ & $l=1$ effective ``range" & -0.34 $\mathrm{fm}^{-1}$ \\

 \hline
\end{tabular}
\caption{Key physical scales in our EFT. The $\be\rightarrow p+\bes$ threshold is $B_{\be}$, the $\bes\rightarrow {}^{3}\mathrm{He}+{}^{4}\mathrm{He}$ threshold is $B_{\bes}$, and the $\bes$ core-excitation energy is $E^{\ast}$.  The $p$-$\bes$ effective mass $\mr\equiv \mn \mc/(\mn+\mc)=7/8 \mn$ and the ${}^{3}\mathrm{He}$-${}^{4}\mathrm{He}$ effective mass $M'_R=12/7 \mn$.  The scattering parameters $a$ and $r$ in the incoming $s$-wave channels are not well determined, but generically obey the hierarchy given here. The numbers for $a_1$ and $r_1$ are extracted from {\it ab initio} ANCs.} \label{tab:dynamicenergyscales}
\end{table}

{\em Energy scales and Lagrangian.---} It is essential to identify all
the pertinent energy scales in this problem
(Table~\ref{tab:dynamicenergyscales}).  The presence of a Coulomb
barrier generates a low-momentum scale, $k_C$, associated with its
height~\cite{KongRavndal,Higa:2008,Ryberg:2013}. $k_C$ is comparable
to the binding momentum of $\be$, $\gamma$, which is our generic
low-energy scale.  The large $p$-$\bes$ scattering lengths yield two
more, quite similar, low-energy scales. All of these generate
non-analytic dependence in the $S$-factor $S(E)$: if $k=\sqrt{2 \mr
  E}$ then $S=S(k_C/k, ka,k/\gamma)$. We keep the full dependence on
these ratios generated by  Halo EFT at LO and so have nontrivial
analytic structure due to long-wavelength properties of Coulomb wave
functions and strong initial-state interactions. This stands in
contrast to the Taylor \cite{WilliamsKoonin,Baye:2000ig, Baye:2000gi,Baye:2005} expansion
sometimes used to describe $S(E)$ around $E=0$. The EFT's high-energy
scale is $\Lambda$, the momentum at which $\bes$ sub-structure is
resolved.  $\Lambda \approx 70$ MeV is set by the threshold for
$\bes\rightarrow {}^{3}\mathrm{He}+{}^{4}\mathrm{He}$ and is the
radius of convergence in the momentum plane of our result for $S(k)$,
which is accurate up to corrections suppressed by
$\gamma/\Lambda$. The $\be$ ground state is bound by $0.1375$ MeV, so
$\gma/{\Lambda} \approx 0.2$.  We use the power counting of
Ref.~\cite{Bd03} to describe this shallow $p$-wave bound state:
$a_{1}^{-1}\sim \Lambda \gamma^{2}$, $r_{1}\sim \Lambda$.  We also
include the core excitation $\besst$ as an explicit degree of
freedom. Its excitation energy $E^*$ is another low-energy scale.

%
%


The Lagrangian we use is $\mathcal{L}_0 + \mathcal{L}_S + \mathcal{L}_{P}$ (cf. Ref.~\cite{Zhang:2013kja}):
\begin{eqnarray}
&&\mathcal{L}_{0}=\fdu{n}{\sigma} \left(i\partial_{t}+\frac{\bigtriangledown^{2}}{2\mn} \right) \fd{n}{\sigma} + \fdu{c}{a}\left(i\partial_{t}+\frac{\bigtriangledown^{2}}{2\mc} \right) \fd{c}{a}\notag \\
&& \quad +  \fdu{d}{\delta}\left(i\partial_{t}+\frac{\bigtriangledown^{2}}{2\mc} \right) \fd{d}{\delta} +\fdu{\pi}{\alpha}\left(i\partial_{t}+\frac{\bigtriangledown^{2}}{2\mnc} + \Delta \right) \fd{\pi}{\alpha} \ , \notag  \\
&&\mathcal{L}_{S}= \gone \fdu{c}{a'} \fdu{n}{\sigma'} \T{a'\sigma'}{i} \T{i}{a \sigma} \fd{c}{a}\fd{n}{\sigma} \notag \\
&& \quad +\gtwo \fdu{c}{a'} \fdu{n}{\sigma'} \T{a'\sigma'}{\alpha} \T{\alpha}{a \sigma} \fd{c}{a}\fd{n}{\sigma} \notag \\ 
&& \quad +\gthree \fdu{d}{\delta}\fdu{n}{\sigma'} \T{\delta \sigma'}{i} \T{i}{a\sigma} \fd{c}{a} \fd{n}{\sigma} + \mathrm{C.C.} \ , \label{eqn:L1} \\
&&\mathcal{L}_{P}=\hone \fdu{\pi}{\alpha}\T{\alpha}{ij}\T{i}{\sigma a} \fd{n}{\sigma} i\left(\V{n}-\V{c}\right)_{j} \fd{c}{a}\notag \\
&&\quad+\htwo \fdu{\pi}{\alpha}\T{\alpha}{\beta j}\T{\beta}{\sigma a} \fd{n}{\sigma} i\left(\V{n}-\V{c}\right)_{j} \fd{c}{a} \notag \\
&& \quad +\hthree \fdu{\pi}{\alpha} \T{\alpha}{jk} \T{k}{\delta \sigma}\fd{n}{\sigma} i(\V{n}-\V{c^{*}})_{j} \fd{d}{\delta} 
+\mathrm{C.C.} \ . \label{eqn:L2}
\end{eqnarray} 
Here $\fd{n}{\sigma}$, $\fd{c}{a}$, $\fd{d}{\delta}$, $\fd{\pi}{\alpha}$ are fields of the proton (``nucleon''), $\bes$ core ($\frac{3}{2}^{-}$), $\besst$ ($\frac{1}{2}^{-}$), and $\be$ ground state ($2^{+}$) respectively; $\gone$ and $\gtwo$ in $\mathcal{L}_S$ are related to ``unnaturally'' enhanced $\sim 1/\gamma$ $s$-wave charged-particle scattering lengths (see Refs.~\cite{KongRavndal,Higa:2008,Ryberg:2013}); $\gthree$ describes $\bes + p \leftrightarrow \besst+p$ and is assumed to be natural, i.e. $\sim 1/\Lambda$~\cite{Zhang:2013kja}; and $h_Y$ are the $p$-wave couplings. The fields' indices are their spin projections with a specific convention:  $\sigma, \delta, \sigma', \delta'=\pm 1/2$, $a,a'=\pm 3/2, \, \pm 1/2$, $\alpha,\beta=\pm 2, \, \pm 1, \, 0$, and $i,j,k=\pm 1, \, 0$;  $\Delta$ is $\fu{\pi}{\alpha}$'s bare binding energy; $\T{...}{...}$s are the C-G coefficients (cf.~\cite{Zhang:2013kja}); $\V{n;c}$ are proton and core velocities.
To implement electromagnetic interactions, we use minimal substitution $\partial_{\mu}\rightarrow \partial_{\mu}+ie Q A_{\mu}$ with $e\equiv \sqrt{4\pi\alpha_{EM}}$ and $Q$ the particle's charge. 

{\em $p$-wave scattering and shallow bound state.---}
We denote $\fu{\pi}{\alpha}$'s dressed propagator as $D_{\alpha}^{\beta}\equiv  D\delta_{\alpha}^{\beta}$. Its self energy receives contributions from $p$-$\bes$ intermediate states in the $\P{3}{2}$ and $\P{5}{2}$ channels, and $p$-$\besst$ in the $\P{3}{2}$ channel \cite{Zhang:2013kja}. In each case the proton and $\bes$ ($\besst$) interact via the Coulomb interaction. We include these Coulomb effects to all orders in $\alpha_{EM}$, thereby extending the calculation of Refs.~\cite{KongRavndal,Higa:2008,Ryberg:2013} to $p$-waves. In practice this involves replacing the plane waves of Ref.~\cite{Zhang:2013kja} by Coulomb wave functions in both external legs and intermediate states. $D(E)$ can then be written in terms of the $a_{1}$ and $r_{1}$ in the $2^{+}$ channel:
\begin{eqnarray}
\frac{-6\pi \mr}{\hpt^{2} \, D} 
&=&-\frac{1}{a_{1}}+\frac{r_{1}}{2} \ke^{2}-2\kc( \ke^{2}+\kc^{2})  H(k_C/k) \notag \\ 
&&-2\kc \frac{\hthree^{2}}{\hpt^{2}} (\kest^{2}+\kc^{2})  H(k_C/\kest) \ . 
\label{eq:D}
\end{eqnarray}
with $H(\eta) =\psi(i\eta)+{1}/{(2i\eta)}-\ln(i\eta)$  \cite{MathHandBook1},  $\hpt^{2}\equiv\hone^{2}+\htwo^{2}$ and $\kest \equiv \sqrt{2\mr (E-E^{\ast})}$.
$a_{1}$ and $r_{1}$ are then functions of the 
Lagrangian parameters $\Delta$, $\hone$, $\htwo$, and $\hthree$. If $k < \gma_{\Delta}$ we recover the standard Coulomb-modified effective-range expansion for $l=1$, i.e. the right-hand side of Eq.~(\ref{eq:D}) becomes $C_{\eta,1}^2 k^3 (\cot \delta_1 - i)$, with
$\delta_1$ the phase shift relative to a Coulomb wave, $C_{\eta,l} \equiv {2^l \exp{(-\pi \eta/2)}|\Gamma(l+1+i \eta)|}/{\Gamma(2l +2)}$, and $\eta \equiv k_C/k$~\cite{Koenig:2012bv,GoldbergerQM}.

We write the $p$-wave Coulomb-distorted $T$-matrix, $T_{CS}$~\cite{Higa:2008,Koenig:2012bv}, in the $p$-$\bes$-$\besst$ Hilbert space  schematically as $V\times D(E) \times V$. ($V$ stands for $n$-$c$($d$)-$\pi$ interactions and $D$ is given by Eq.~(\ref{eq:D}).)  Since $T_{CS}$ has a  pole at $B_{\be}$ we have $D^{-1}(\ke=i\gma)=0$. The residue then yields a relation for the wave-function renormalization factor, $Z$:
\begin{widetext}
\begin{align}
\frac{6\pi}{Z}+\hpt^{2} \ r_{1}=2\frac{\kc}{\gamma} \bigg\{\frac{\hpt^{2}}{\gamma^2} \bigg[2 \gamma^3 \tilde{H}\left(\frac{k_C}{\gamma}\right)+(k_C^3-k_C \gamma^2)\tilde{H}^{\,\bold{'}}\left(\frac{k_C}{\gamma}\right)\bigg] +\frac{\hthree^{2}}{{\gma^\ast}^2} \bigg[2 {\gamma^\ast}^3 \tilde{H}\left(\frac{k_C}{\gamma^*}\right)+(k_C^{3}-k_C {\gamma^{\ast}}^2)\tilde{H}^{\,\bold{'}}\left(\frac{k_C}{\gamma_*}\right)\bigg] \bigg\},
\label{eqn:Zexpr}
\end{align}
\end{widetext}
where, for convenience, we define $\tilde{H}(z)\equiv H(-iz)$, $\tilde{H}^{\,\bold{'}}(z) \equiv {d\tilde{H}(z)}/{dz}$. 
The $\be\rightarrow p+\bes (\besst)$ ANCs, $C_{(\P{5}{2})}$, $C_{(\P{3}{2})}$, and $C_{(\P{3}{2}^*)}$  are then given by (cf. Ref.~\cite{Zhang:2013kja}):
\begin{eqnarray}
\frac{C^{2}_{(Y)}}{h_{Y}^{2}\gma^{2} \Gamma^2(2+k_C/\gamma)} =\frac{C^{2}_{(\P{3}{2}^{\ast})}}{\hthree^{2}\gma^{\ast 2} \Gamma^2(2+k_C/\gamma^\ast)} =\frac{Z}{3\pi},\ \label{eqn:ancrone}
\end{eqnarray}
with $Y=\P{3}{2}$ and $\P{5}{2}$.

\begin{figure} [h!]
\centering
\includegraphics[width=8.3cm, angle=0]{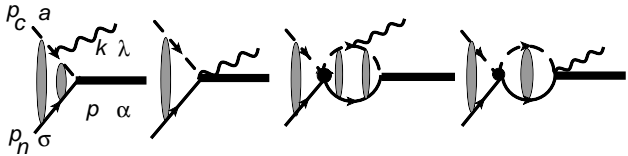}
\caption{Diagrams for capture. The line assignments are designated by the spin indices; the shaded blobs denote the full Coulomb Green functions; the bold vertex in the last two diagrams means $S$-wave multiple scattering with the Coulomb interaction included to all orders in $\alpha_{EM}$~\cite{KongRavndal,Zhang:2013kja}. Diagrams with the photon coupled to the proton are not shown, but are included in our calculation.} \label{fig:ncapturediag}
\end{figure}

{\em Leading-order proton capture.---}
The relevant diagrams are shown in Fig.~\ref{fig:ncapturediag}.  Let us first focus on initial total spin $S_{i}=1$ and use the notation,  $\langle\pi^{\alpha}|L_{EM}|\cw{\vec{p}}{+}, \delta, a\rangle \equiv \T{i}{\delta a} \T{\alpha}{ij} \mathcal{M}_{j}$.  The LO $\mathcal{M}_{j}$ is computed in coordinate space, and decomposed into  contributions from incoming $s$- and $d$-waves.
\begin{widetext}
\begin{align}
 \mathcal{M}_{j} &=(-i)  C_{\eta,0} C_{(\P{3}{2})}^{\mathrm{LO}}\frac{Z_{eff}}{\mr} \frac{2\pi}{\sqrt{3}} \left(\gma^{2}+\ke^{2}\right)  \left[e^{i\sigma_{0}}\es_{j}\dY{0}{0}(\vecpt{p})\mathcal{S}(\S{3}{1})+e^{i\sigma_{2}}\es_{k} \sqrt{2} \T{j}{ka} \dY{2}{a}(\vecpt{p}) \mathcal{D} \right]  \  , \notag \\
\mathcal{S}(X) \equiv \int_{0}^{+\infty} dr  & W_{-\eta_{B},\frac{3}{2}}(2\gma r) r  \left[\frac{ C_{\eta,0}\cwg{0}(\ke,r)}{-a_{(X)}^{-1}-2\kc H(\eta)} +\frac{\cwf{0}(\ke,r)}{C_{\eta,0} \ke} \frac{-a_{(X)}^{-1}-2\kc \mbox{Re}\left[ H(\eta)\right]}{-a_{(X)}^{-1}-2\kc H(\eta)}\right] ,\notag\\
 \mathcal{D} &\equiv \int_{0}^{+\infty} dr  {W_{-\eta_{B},\frac{3}{2}}(2\gma r)}   r  \frac{\cwf{2}(\ke,r)}{C_{\eta,0} \ke} \ . \label{eqn:SDdef}
\end{align}
Here $C_{(\P{3}{2})}^{\mathrm{LO}}$ is the LO ANC (see below);
${Z_{eff}}/{\mr}\equiv {e Q_{n}}/{\mn}-{e Q_{c}}/{\mc}$; $\es$ is the
photon polarization vector; $\sigma_{l}$ is the Coulomb phase shift;
$F_{l}$ and $G_{l}$ are Coulomb wave functions for angular momentum
$l$, and $W_{-\eta_{B},\frac{3}{2}}$ is a Whittaker function with
$\eta_B \equiv k_C/\gamma$ \cite{GoldbergerQM}. Proceeding similarly
with $S_{i}=2$ we obtain the same result but with
$(C_{(\P{3}{2})}^{\mathrm{LO}},\, \aone)\rightarrow
(C_{(\P{5}{2})}^{\mathrm{LO}},\, \atwo )$. The $s$-wave scattering
lengths, $\aone$ and $\atwo$, describe LO incoming channel multiple
scattering (ICMS) effects.  We thus obtain the $S$-factor including
all initial channels:
\begin{eqnarray}
S(E)&=&
\frac{e^{2\pi \eta}}{e^{2\pi\eta}-1} \frac{Z_{eff}^{2}}{\mr^{2}}  \frac{\pi}{24} \omega \kc\left(\gma^{2}+\ke^{2}\right)^{2} \  \frac{5}{3} \bigg[ {C_{(\P{3}{2})}^{\mathrm{LO}}}^{2}\left(\mid \mathcal{S}(\S{3}{1})\mid^{2}+2 \mid \mathcal{D}\mid^{2}\right)+{C_{(\P{5}{2})}^{\mathrm{LO}}}^{2}\left(\mid \mathcal{S}(\S{5}{2})\mid^{2}+2 \mid \mathcal{D}\mid^{2}\right) \bigg] \ ,
\label{eq:SE}
\end{eqnarray}

\end{widetext}

\begin{table}
   \centering
   \begin{tabular}{|c|c|c|c|c|} \hline
           & $C_{(\P{3}{2})}$
           & $C_{(\P{5}{2})}$   
           & $\aone $
           & $\atwo $ \\  \hline
           Nollett 
           & $-0.315(19)$
           & $-0.662(19)$ 
           & $$
           & $$  \\  \hline
    Navratil 
           & $-0.294$
           & $-0.650$
           & $-5.2$
           & $-15.3$ \\   \hline      
 Tabacaru 
           & $ 0.294(45)$
           & $0.615(45)$
           & $$ 
           & $$ \\  \hline  
  Angulo 
           & $$
           & $$
           & $25(9)$
           & $-7(3)$     \\  \hline 
   \end{tabular}
   \caption{ 
Input parameters.  ``Nollett'' \cite{Nollett:2011qf} and ``Navratil'' \cite{Navratil:2011sa}  are  {\it ab initio}  results, while ``Tabacaru'' \cite{Tabacaru:2005hv} and ``Angulo'' \cite{Angulo:2003} are from experiment. The units are $\mathrm{fm}^{-1/2}$ and $\mathrm{fm}$ for ANCs and scattering lengths respectively.} \label{tab:ancanda}
\end{table}

\begin{table}
   \centering
   \begin{tabular}{|c|c|c|c|c|c|c|} \hline
           & $S(0)$ (eV b)
           & $S_{(\S{3}{1})}(0)$
           & $d_{1}$($\mathrm{MeV}^{-1}$)  
           & $d_{2}$ ($\mathrm{MeV}^{-2}$) \\  \hline
           No+A 
           & $18.2 \pm 1.2$
           & $3.1 \pm 0.4$ 
           & $-1.62$ 
           & $10.3$ \\  \hline
           Na 
           & $17.8$
           & $3.0$
           & $-1.26$
           & $10.8$ \\   \hline      
           T+A 
           & $15.7 \pm 2.7$
           & $2.7 \pm 0.8$
           & $-1.62$ 
           & $10.3$ \\  \hline  
           Ref.~\cite{Adelberger:2010qa}
           & $20.8 \pm 1.6$
           & 
           &$-1.5 \pm 0.1$
           & $6.5 \pm 2.0$\\ \hline
   \end{tabular}
   \caption{Results for Taylor-expansion coefficients. No+A, Na, and T+A, are the results from using  ``Nollett''+``Angulo'', ``Navratil'', and ``Tabacaru''+``Angulo'' inputs listed in Table~\ref{tab:ancanda}.  The uncertainties in No+A and T+A entries are due to ANC uncertainties listed in Table~\ref{tab:ancanda}.  Recommended values from Ref.~\cite{Adelberger:2010qa} are shown for comparison.} \label{tab:results}
\end{table}

\begin{figure}
\centering
\includegraphics[width=6cm, angle=-90]{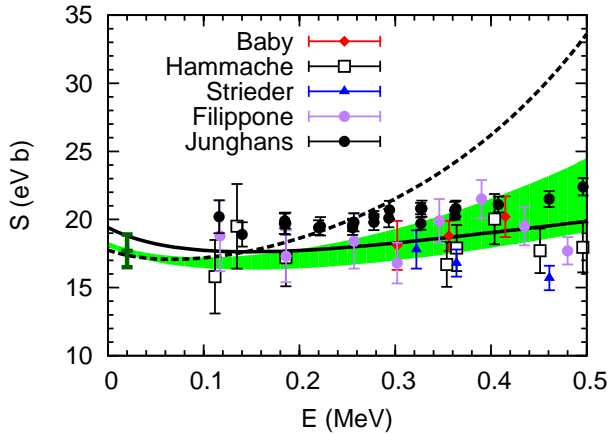}
\caption{$E1$ $S$-factor for proton capture on $\bes$. 
The green band is the range of $S(E)$ in the LO calculation with ``No + A" input (cf.~Table~\ref{tab:results}), varying  scattering lengths within their $1\sigma$ errors. The  (nearly energy-independent) impact of the ANC error bar on the prediction is indicated by the error bar on the ``No + A" band at 20 keV.
The data are from Refs.~\cite{Filippone:1984us,Baby:2002hj,Junghans:2003bd,Junghans:2010zz}.  The dashed line is the LO result with ``Na" input. The solid line shows the corresponding full RGM-NCSM result~\cite{Navratil:2011sa} for comparison. } \label{fig:results1}
\end{figure}

{\em Results.---}
 Table~\ref{tab:ancanda} collects information on $\be\rightarrow
 \bes\,+\,p$ ANCs and $p$-$\bes$ scattering lengths.  The \textit{ab
   initio} ANCs of Nollett \& Wiringa were computed using variational Monte Carlo
 with the Argonne $v18$ + Urbana IX
 Hamiltonian~\cite{Wiringa:1995,Pudliner:1995}, which includes
 three-nucleon terms.  They are consistent with those from
 experiment~\cite{Tabacaru:2005hv} and may have smaller errors (though
 errors reported for the calculation account only for Monte Carlo
 sampling).  These ANCs are also consistent with \textit{ab initio}
 ANCs found from an SRG-N$^2$LO two-body interaction in the RGM-NCSM
 approach by Navratil \textit{et al.}~\cite{Navratil:2011sa}.  We
 extended the calculations in Ref.~\cite{Nollett:2011qf} to include
 the $\be\rightarrow \besst\,+p$ ANC, $C_{(\P{3}{2}^{\ast})}=-0.3485(51)$
 $\mathrm{fm}^{-1/2}$.

Using the Nollett ANCs, the ${}^8$B binding energy, and
Eqs.~(\ref{eqn:ancrone}) and (\ref{eqn:Zexpr}) we obtain the results
for $a_1$ and $r_1$ shown in Table~\ref{tab:dynamicenergyscales} --
values that conform to the scaling assigned to $a_1$ and $r_1$. Given
this input information, we work strictly to LO, so the right-hand side
of Eq.~(\ref{eqn:Zexpr}) is set to zero, yielding
$Z^{\mathrm{LO}}={-6\pi}/{\hpt^{2}r_{1}}$. We then define
corresponding ``LO ANCs" $C^{\mathrm{LO}}_{(\cdots)}$ via
Eq.~(\ref{eqn:ancrone}), which differ from the full ANCs by
${C^{\mathrm{LO}}_{(\cdots)}}^{2}/C^{2}_{(\cdots)}=Z^{\mathrm{LO}}/Z=0.87$. From
these and Eq.~(\ref{eq:SE}) we compute the LO E1 $S$-factor for
${}^7\mathrm{Be}(p,\gamma) {}^8 \mathrm{B}$.  At low energies, $E <
0.1$ MeV, a quadratic approximation $S(E)=S(0)(1+d_{1}E+d_{2}E^{2})$
is customary~\cite{Baye:2000ig,
  Baye:2000gi,Baye:2005}. Table~\ref{tab:results} lists results, based
on different inputs, for the full $S(0)$, the $S_i$=1 channel
contribution $S_{(\S{3}{1})}(0)$, $d_1$, and $d_2$.  The peripheral
nature of the capture in this regime means both the $S_i=1$ and
$S_i=2$ contributions to $S(0)$ depend only very weakly on ICMS---as
seen in Refs.~\cite{Baye:2000ig, Baye:2000gi,Baye:2005}. As a result,
$S_{(\S{3}{1})}(0)$ and $S_{(\S{5}{2})}(0)$ essentially both scale
with their corresponding LO ANCs. The recommended value of
Ref.~\cite{Adelberger:2010qa} for $S(0)$, which was fitted to direct
capture data, is consistent with our $S(0)$ within uncertainties
(including the higher-order EFT uncertainty of $\approx 10$\%, see
below).  The $d_1$ and $d_2$ of Ref.~\cite{Adelberger:2010qa} were
obtained by fitting quadratics to model curves (not data) in the range
$0\leq E\leq 50$ keV; we define our coefficients in the same way and
find consistent results, though our $d_2$ are near the high end of a
large variation among models.  Exact derivatives computed at threshold
are in close agreement with potential models~\cite{Jennings:1998ky}.

In Fig.~\ref{fig:results1} we show the LO Halo EFT prediction for
$S(E)$ obtained from the ``Nollett" ANCs and the $1\sigma$ range of
experimental scattering lengths~\cite{Angulo:2003}. The errors on the
ANCs smear out the prediction by an additional $\sim 10\%$ at all
energies as shown by an error bar at 20 keV.  In contrast, the impact
of scattering length uncertainties increases dramatically with energy:
when $E<0.1$ MeV, it is $< 5\%$, but it
reaches 20\% at $E \approx 0.4$ MeV. The consequent range of Halo-EFT
predictions is consistent with direct capture data for $0.1\leq E \leq
0.4$ MeV (below the $M1$ resonance) and with indirect $S$-factors
derived from Coulomb breakup
\cite{Davids:2003aw,Kikuchi:1998,Schumann:2005fk}.  However, the
energy dependence seen at the upper edge of the band in
Fig.~\ref{fig:results1} is not consistent with the trend of the data.
This rapidly rising $S(E)$ happens because the real part of the
denominator in Eq.~(\ref{eqn:SDdef}) vanishes at some energy $E_p > 0$
if $\atwo < 0$. $E_p$ is quite low with $\atwo=-15.3$ fm from
RGM-NCSM~\cite{Navratil:2011sa}, so the dashed Halo-EFT curve that we
obtain with the RGM-NCSM inputs rises significantly faster than the
data.  We emphasize that this increase is {\it not} the well-known $M1$
resonance contribution~\cite{Navratil:2011sa,Adelberger:2010qa}, since
the resonant channel is not included in our calculation.  We show
$S(E)$ up to $E=0.5$ MeV, where the resonance has little impact and
ICMS effects can generate marked energy dependence in $S(E)$.

The energy dependence of the full RGM-NCSM calculation (solid curve) is
much gentler than the Halo EFT result using its \atwo, and the
difference must reflect higher-order terms in the EFT capture amplitude.
We assess the impact of next-to-leading-order (NLO) contributions in
two ways. First, we introduce an effective-range term in the $s$-wave
scattering amplitude. For simplicity we choose the same effective
range, $r_0$, for both channels.  Varying $r_0$ from its LO value of 0
fm to 2 fm changes the $S$-factor by less than 10\%, even when
$\atwo=-15.3$ fm and $E=0.4$ MeV so that the effect is largest.
Second, we change the lower limit of integration, $r_{\mathrm{min}}$,
in Eq.~(\ref{eqn:SDdef}) to estimate the size of higher-order,
short-distance effects like the contact operators that enter the
capture amplitudes at NLO~\cite{Zhang:2013kja}. With ``Na" input and
$r_0=2$ fm, an $r_{\mathrm{min}}$ of 2 fm flattens $S(E)$, reducing
it by $10\%$ at 0.4 MeV.  Varying parameters by hand, we find
that $r_0=r_{\mathrm{min}}=3$ fm brings the energy dependence of our LO ``Na'' $S(E)$ curve
close to that of the full RGM-NCSM result.

We emphasize that although the precise magnitude and shape of $S(E)$
predicted by Halo EFT will be modified by higher-order contributions,
the sensitivity to $p$-${}^7$Be scattering parameters we have diagnosed here
should persist.  We conclude that it is important
to improve the accuracy of $s$-wave scattering length (and perhaps
effective range) measurements in order to constrain the extrapolation
of $S(E)$ data to zero energy. This supports similar conclusions from
other formalisms \cite{Davids:2003aw,Adelberger:2010qa,Descouvemont:2004}.
Models with incorrect scattering lengths get a key non-analyticity of
the low-energy capture amplitude wrong.

{\em Summary.---}
We have applied Halo EFT to radiative proton capture on $\bes$. In Halo EFT the short-distance piece of the Coulomb-nuclear interference is entangled with the pure nuclear amplitude, so this process is not straightforwardly connected to radiative neutron capture on ${}^{7}\mathrm{Li}$, although isospin symmetry is regularly used to relate the two in models. Discussion of this issue and other details of our calculation will be presented elsewhere~\cite{Zhang:2014}. The strategy used here is however the same as in Ref.~\cite{Zhang:2013kja}: fix the EFT couplings using  {\it ab initio} ANCs and experimental binding energies and scattering lengths.

Our LO Halo EFT result for $S(0)$ is $\approx 10\%$ below the
recommended value \cite{Adelberger:2010qa}, but is consistent with it
within the combined EFT and ANC uncertainties. The significant
uncertainties in the experimental $a_{(\S{5}{2})}$ and
$a_{(\S{3}{1})}$ mean that the energy dependence of $S(E)$ is not well
constrained above 0.1 MeV, although the central values produce a trend
that agrees with direct capture data quite well.  This shows the
importance of improved measurements of $p$-${}^7$Be scattering, which
would render the extrapolation of capture data to solar energies more
reliable. On the theory side a next-to-leading-order calculation of $p
+ {}^7 \mathrm{Be} \rightarrow {}^8 \mathrm{B} + \gamma$ will be
important to reduce the EFT uncertainty and so make Halo EFT
competitive as a tool for determining this key input to predictions of
solar neutrino fluxes.

{\em Acknowledgments.---}
X.Z.~and D.R.P.~acknowledge support from the US Department of Energy under grant DE-FG02-93ER-40756. K.M.N. acknowledges support from the Institute of Nuclear and Particle Physics at Ohio University. We thank Carl Brune for useful comments on the manuscript.

\bibliographystyle{apsrev}

\end{document}